\documentclass[final,5p,times,number]{elsarticle}

\usepackage[T1]{fontenc}
\usepackage[utf8]{inputenc}
\usepackage[english]{babel}
\usepackage{amsmath}
\usepackage{amssymb}
\usepackage{graphicx}
\usepackage[bookmarks=true,bookmarksnumbered=false,bookmarksopen=false,
 breaklinks=true,pdfborder={0 0 0},backref=false,colorlinks=false]{hyperref}

\journal{Physics Letters A}

\begin{document}

\begin{frontmatter}

\title{Weak irreducibility as a spectral criterion for phase coexistence}

\author{Onofre Rojas}
\address{Department of Physics, Institute of Natural Science, Federal University of Lavras, Lavras, MG, Brazil}

\begin{abstract}
Conventional finite-size theories describe first-order coexistence 
through free-energy competition and interfacial tunneling, but
the distinct spectral roles of sector balance and inter-sector 
connectivity are not usually explicit. We show that pseudo-transitions 
and thermodynamic phase coexistence are governed by two spectral 
coordinates: sector imbalance locates balance, while connectivity 
determines whether that balance remains avoided. At balance, finite 
connectivity produces a unique dominant state with equal spectral 
weights in the symmetrized two-sector representation. For short-range 
systems with positive interface tension, interfacial costs suppress 
the connectivity exponentially, close the coexistence splitting, and 
asymptotically restore reducibility. Thus, genuine phase coexistence 
emerges as the singular limit of finite-size avoided coexistence, 
whereas one-dimensional pseudo-transitions retain finite connectivity. 
In a decorated bilayer Ising model, the coexistence gap at 
independently established balance isolates the normalized 
inter-sector connectivity and recovers the interface tension 
without imposing an interface.
\end{abstract}


\end{frontmatter}

\section{Introduction}

Phase coexistence is one of the central phenomena of statistical mechanics.
In systems undergoing first-order phase transitions, competing thermodynamic
phases become degenerate and coexist at equal free energy. The resulting
coexistence regime governs nucleation, metastability, interfacial
phenomena, and finite-size effects \citep{Pirogov76,Peierls36,Borgs,Binder87,Lee}.
Despite its fundamental importance, the microscopic origin of coexistence
is usually formulated in geometric or thermodynamic terms, whereas
its spectral structure remains less directly understood.
Restricted-ensemble constructions, finite-size scaling, and
partition-function zeros each capture important aspects of the same
phenomenon: the separation of competing phases, the rounding induced
by finite systems, and the emergence of a thermodynamic singularity.
What is still missing is a compact operator-level criterion that says
when a finite-size spectral splitting is merely an avoided crossing
and when it is the finite-size remnant of a reducible coexistence
manifold.

Restricted-phase and contour methods describe phase separation and
interfacial suppression~\citep{Pirogov76,Pirogov75,bricmont,Friedli},
while finite-size treatments emphasize interfacial barriers and
exponentially weak phase mixing~\citep{Borgs,Binder87,Lee,Binder84,biskup}.
Closely related nearly disconnected low-energy sectors also occur in
decorated, frustrated, and pseudo-transition models~\citep{pseudo,ky-lin,ninio,cuesta,Lavis-2015,unv-cr-exp,panov2021,finite-chain,ph-bd,phya-o25,yasinskaya,katarina,Yasinskaya24,Braz25,Sznajd24,Sznajd25,Jung,chapman2024}.

From the transfer-matrix perspective, equilibrium states are described
in the spectrum of a positive operator. For finite systems, the Perron-Frobenius
theorem \citep{ky-lin,ninio} guarantees a unique dominant eigenvalue
whenever the transfer operator remains irreducible. Consequently,
exact level crossings and finite-temperature nonanalyticities are
excluded in one-dimensional systems with short-range interactions
\citep{cuesta,Lavis-2015}.
Nevertheless, many statistical-mechanical systems exhibit competing
low-energy structures whose thermodynamic signatures closely resemble
phase coexistence \citep{panov2021,yasinskaya,katarina}.

This raises a question not answered by conventional interfacial
descriptions: which spectral property distinguishes an avoided
coexistence crossover from genuine thermodynamic coexistence?

In plain terms, a small gap tells us that two sectors are close, but not
whether they remain connected in the thermodynamic limit.

We answer it through weak irreducibility: the transfer operator remains
globally irreducible, but competing low-energy sectors communicate
only through a small connectivity. Unlike a description based solely
on free-energy degeneracy, this construction separates two independent
ingredients, sector balance and sector communication, and tracks
both the eigenvalues and the dominant state. A small gap alone is
insufficient; at independently established balance it directly probes
the residual communication.

In one-dimensional systems, finite inter-sector connectivity prevents exact spectral
crossings and gives rise to pseudo-transition behavior
\citep{ph-bd,phya-o25}. Recent studies
span decorated Ising ladders and chains
\citep{chapman2024,w-yin-prl,w-yin-prr,strecka-dmd},
mixed-spin systems \citep{katarina,Rodrigo}, and other frustrated or
Potts-like models \citep{panov2021,yasinskaya,Yasinskaya24,Braz25}.
Different interpretations have been proposed, including quasiphases
\citep{pseudo,unv-cr-exp}, entropy competition, avoided level crossings,
and connections with Widom-line physics \citep{chapman-26}. Recent
work has begun connecting these crossovers to higher-dimensional
first-order behavior \citep{Sznajd24,Sznajd25,Jung,chapman-26,MataragkasTM25}.
Previous spectral studies identified nearly reducible one-dimensional
transfer matrices \citep{phya-o25} or extracted interfacial information
from exponentially closing gaps \citep{Jung,MataragkasTM25}. The
advance here is not merely another small-gap diagnostic: balance and
connectivity are independently determined, providing an operational
distinction between an avoided crossing and a reducible coexistence
manifold.

Our central result is simple: balance locates coexistence, whereas the
asymptotic fate of connectivity classifies it. Finite connectivity
produces avoided coexistence and pseudo-transition behavior; vanishing
connectivity restores a reducible manifold of thermodynamic phases.
The dimensionless spectral gap measures this communication, while
the dominant-state weights diagnose sector balance.
Thus interfacial tunneling is not only a qualitative mechanism for
finite-size rounding; at balance, it becomes the measurable
connectivity that controls the approach to reducibility. This makes
the criterion useful for exact transfer matrices and numerical
transfer operators, where one can test coexistence without explicitly
constructing an interface.

The paper is organized as follows. Section~\ref{sec:SM1} develops the
effective two-sector transfer-operator reduction, introduces the imbalance
and connectivity coordinates, and uses their asymptotic behavior to
distinguish pseudo-transitions from thermodynamic coexistence. Section~\ref{sec:SM4}
derives the local thermodynamic consequences of avoided coexistence,
including the rounded coexistence width, response enhancement, and spectral
length scale. Section~\ref{sec:SM5} applies the framework to a decorated
bilayer Ising model, where the balance condition is known independently and
the coexistence gap recovers the exact Ising interface tension without
imposing an interfacial configuration. The final section discusses the
relation with restricted ensembles, finite-size scaling, and partition-function
zeros, and summarizes the resulting spectral criterion. An appendix records
the balanced-branch entropy-density construction and clarifies its order of
limits.

\section{Effective two-sector spectral reduction and asymptotic reducibility}\label{sec:SM1}
Here we formulate an operational spectral description of
coexistence based on competing low-energy sectors and weak inter-sector
connectivity. Sector imbalance locates balance, whereas connectivity
controls the hybridization of the competing sectors within the effective 
coexistence manifold. This separation is essential: a small spectral gap 
alone does not diagnose coexistence; evaluated at independently established 
sector balance, it directly probes the residual communication between sectors.

\subsection{Weakly reducible transfer operators}

We consider a $d$-dimensional system in a cylindrical geometry with
longitudinal length $N$ and transverse linear size $L$, so that
the total volume is $NL^{d-1}$. The transfer operator acts along
the longitudinal direction. Throughout this work, the thermodynamic
limit $N\rightarrow\infty$ is taken from the outset, while the transverse
scale $L$ remains finite. For $d=1$, one has $L^{d-1}=1$, whereas
for $d\geqslant2$ all intensive thermodynamic quantities are understood
per transverse volume $L^{d-1}$. The partition function can therefore
be written as
\begin{equation}
Z_{N,L}={\rm Tr}\left(\mathcal{V}^{N}_{L}\right),
\end{equation}
where $\mathcal{V}_{L}$ denotes the transfer operator associated
with a transverse section of size $L^{d-1}$.

The objective is to identify the amplitudes governing sector 
communication and give interfacial tunneling a measurable 
operator-level meaning.

\subsubsection{Scope and assumptions}

The classification below requires a finite set of dominant low-energy
sectors separated from the remaining spectrum. In the two-sector case
treated explicitly, the finite-$L$ transfer operator is non-negative
and irreducible, while its restricted blocks possess isolated dominant
eigenvalues. Inter-sector communication is assumed to be generated by
the off-diagonal blocks, with no lower-free-energy process bypassing
the minimal connection. The interfacial scaling further requires
short-range interactions and positive interface tension. Finally, a
temperature-driven first-order transition requires a transverse
crossing of the restricted free-energy branches; balance without this
condition need not imply a latent-heat discontinuity. These hypotheses
separate the general spectral classification from the additional
conditions needed for its interfacial realization.

\subsubsection{Restricted sectors and transfer-operator decomposition}

Near a coexistence regime between two competing thermodynamic phases,
or their pseudo-transition counterparts, the configuration space naturally
separates into two restricted sectors, denoted by $a$ and $b$.
The labels identify the two competing phase or low-energy subspaces; in
the decorated bilayer they correspond to the FM-like and FR-like sectors.
Configurations are connected by typical fluctuations within a sector,
whereas changing sector requires the process represented by an
off-diagonal block.

In the strictly reducible limit, trajectories connecting configurations
belonging to different sectors are absent. The transfer operator then
decomposes into independent restricted operators,
\begin{equation}
\mathcal{V}_{L}=\mathbf{V}^{(L)}_{a}\oplus\mathbf{V}^{(L)}_{b}.
\end{equation}

For finite transverse width, however, residual processes generally induce weak
communication between the sectors. The transfer operator can then
be written in block form as
\begin{equation}
{\bf \mathcal{V}}_{L}=\begin{pmatrix}\mathbf{V}^{(L)}_{a} & \mathbf{W}^{(L)}_{ab}\\
\mathbf{W}^{(L)}_{ba} & \mathbf{V}^{(L)}_{b}
\end{pmatrix},\label{eq:VL}
\end{equation}
where $\mathbf{V}^{(L)}_{a}$ and $\mathbf{V}^{(L)}_{b}$ describe
the restricted sector dynamics, while $\mathbf{W}^{(L)}_{ab}$ and
$\mathbf{W}^{(L)}_{ba}$ generate transitions between sectors.
With the row-column convention of Eq.~(\ref{eq:VL}),
$\mathbf W^{(L)}_{ab}:\mathcal H_b\to\mathcal H_a$ maps sector $b$ to
sector $a$, whereas $\mathbf W^{(L)}_{ba}:\mathcal H_a\to\mathcal H_b$
maps sector $a$ to sector $b$; the two blocks need not be equal. The
reducible limit of Eq.~(\ref{eq:VL}) is recovered when
$\mathbf{W}^{(L)}_{ab}=\mathbf{W}^{(L)}_{ba}=0$.

Let
\begin{equation}
\mathbf{V}^{(L)}_{a}|r^{(L)}_{a}\rangle=v^{(L)}_{a}|r^{(L)}_{a}\rangle,\qquad\mathbf{V}^{(L)}_{b}|r^{(L)}_{b}\rangle=v^{(L)}_{b}|r^{(L)}_{b}\rangle,
\end{equation}
be the dominant eigenstates of the restricted transfer operators.
Introducing the corresponding left eigenvectors,
\begin{equation}
\langle l^{(L)}_{\alpha}|r^{(L)}_{\gamma}\rangle=\delta_{\alpha\gamma},\quad\alpha,\gamma\in\{a,b\},
\end{equation}
the effective transition amplitudes between the dominant sector states
are defined by
\begin{equation}
w^{(L)}_{ab}=\langle l^{(L)}_{a}|\mathbf{W}^{(L)}_{ab}|r^{(L)}_{b}\rangle,\qquad w^{(L)}_{ba}=\langle l^{(L)}_{b}|\mathbf{W}^{(L)}_{ba}|r^{(L)}_{a}\rangle.\label{eq:wab-wba}
\end{equation}

These amplitudes quantify the residual communication between the dominant
states of the competing sectors. In the reducible limit they vanish
identically, whereas for finite transverse width they remain nonzero
and act as an inter-sector coupling.

\subsubsection{Inter-sector connectivity and interfacial suppression}\label{subsec:sm1A2}

For finite-size first-order coexistence in dimensions $d\geqslant2$
with positive interface tension,
the restricted sectors can be identified with the pure phases constructed
within Pirogov-Sinai theory~\cite{Pirogov76,Pirogov75,bricmont,Friedli}.
Related finite-size descriptions of first-order transitions likewise
emphasize restricted phases, interfacial barriers, and exponentially
small phase mixing~\cite{Borgs,Binder87,Lee,Binder84,biskup}. In this framework,
the partition function admits a contour representation in which fluctuations
around a given phase are described by interfaces and contours.

The fundamental ingredient is the Peierls estimate~\cite{Peierls36},
\begin{equation}
z(\Gamma)\leqslant C\,e^{-\beta\tau|\Gamma|},
\end{equation}
where $z(\Gamma)$ denotes the statistical weight of a contour $\Gamma$,
$\tau>0$ is the interfacial free energy, and $|\Gamma|$ is the interface
area. Here $\beta=(k_{\rm B}T)^{-1}$ denotes the inverse temperature.

A transition between sectors $a$ and $b$ requires a contour spanning
the entire transverse section of the cylinder. The minimal spanning
contour therefore satisfies
\begin{equation}
|\Gamma_{{\rm min}}|=L^{d-1},
\end{equation}
up to geometry-dependent constants absorbed into $\tau$.

Assuming that no lower-cost process connects the two restricted
sectors, the minimal spanning contour fixes the leading exponential
behavior of the inter-sector amplitudes,
\begin{alignat}{1}
w^{(L)}_{ab}= & A_{ab}(L,\beta)\,e^{-\beta\tau L^{d-1}},\label{eq:wab}\\
w^{(L)}_{ba}= & A_{ba}(L,\beta)\,e^{-\beta\tau L^{d-1}},\label{eq:wba}
\end{alignat}
where the prefactors $A_{ab}$ and $A_{ba}$ contain geometry-dependent
subexponential corrections. Thus, the Peierls estimate establishes
the exponential scale; it does not by itself fix these prefactors.
Since the transfer operator is constructed from non-negative
Boltzmann weights, $w^{(L)}_{ab}$ and $w^{(L)}_{ba}$ are non-negative
and become strictly positive whenever the sectors communicate.

Therefore, communication between competing phases becomes exponentially
suppressed as the transverse size increases. This suppression reflects
the increasing free-energy cost associated with configurations containing
interfaces spanning the transverse section.

For finite-state, short-range one-dimensional systems whose
finite-temperature transfer operator is irreducible, the interpretation
in terms of macroscopic interfaces is absent. We drop the superscript
$L$ and write $w_{ab}$ and $w_{ba}$, which remain well defined through
Eq.~(\ref{eq:wab-wba}). Their microscopic origin is weak communication
between competing low-energy sectors, and at any finite balance
temperature the effective connectivity is nonzero, although it may be
exponentially small in microscopic parameters. Singular zero-temperature
limits, reducible constraints, long-range interactions, and infinite
local state spaces lie outside this statement.

The coexistence problem is therefore reduced to understanding how
the inter-sector amplitudes $w^{(L)}_{ab}$ and $w^{(L)}_{ba}$ influence
the dominant restricted-sector eigenvalues $v^{(L)}_{a}$ and $v^{(L)}_{b}$.
These quantities constitute the fundamental building blocks of the 
finite-size coexistence problem. Their asymptotic role in distinguishing 
avoided coexistence from thermodynamic coexistence is developed below.

\subsection{Low-energy coexistence description}

The amplitudes $w^{(L)}_{ab}$ and $w^{(L)}_{ba}$ provide the mechanism
through which competing sectors communicate. We now show how these
amplitudes determine both the low-lying coexistence spectrum and the
structure of the dominant eigenstates.

\subsubsection{Effective coexistence manifold}\label{subsec:SM1B1}

Near coexistence, we assume that two restricted-sector states become
nearly degenerate and remain spectrally separated from higher modes.
The corresponding
coexistence manifold is therefore defined as $\mathcal{M}^{(L)}_{c}={\rm span}\left\{ |a_{L}\rangle,|b_{L}\rangle\right\} $,
where $|a_{L}\rangle\equiv|r^{(L)}_{a}\rangle$, $|b_{L}\rangle\equiv|r^{(L)}_{b}\rangle$.

The projection operator onto this manifold is
\begin{equation}
\mathbf{P}^{(L)}_{c}=|a_{L}\rangle\langle l^{(L)}_{a}|+|b_{L}\rangle\langle l^{(L)}_{b}|.
\end{equation}
This is a biorthogonal projector:
\begin{equation}
\left(\mathbf{P}^{(L)}_{c}\right)^2=\mathbf{P}^{(L)}_{c},
\end{equation}
but, for a nonsymmetric transfer operator, it is generally not an
orthogonal projector,
\begin{equation}
\left(\mathbf{P}^{(L)}_{c}\right)^\dagger\neq \mathbf{P}^{(L)}_{c}.
\end{equation}

In strictly one-dimensional systems, the transverse section is trivial
and one may simply write $|a\rangle$ and $|b\rangle$. We retain
the notation $|a_{L}\rangle$ and $|b_{L}\rangle$ because it remains
valid for arbitrary spatial dimension.

At leading order in the separation from the remaining spectrum,
projecting the full transfer operator onto $\mathcal{M}^{(L)}_{c}$
yields
\begin{equation}
\mathcal{V}^{(L)}_{{\rm eff}}=\mathbf{P}^{(L)}_{c}\mathcal{V}_{L}\mathbf{P}^{(L)}_{c}=\begin{pmatrix}v^{(L)}_{a} & w^{(L)}_{ab}\\
w^{(L)}_{ba} & v^{(L)}_{b}
\end{pmatrix}.
\end{equation}
More generally, elimination of the complementary subspace
$\mathbf Q_c^{(L)}=1-\mathbf P_c^{(L)}$ generates a Schur-complement
correction of the form
\begin{equation}
\mathbf P_c^{(L)}\mathcal V_L\mathbf Q_c^{(L)}
\left(\lambda-\mathbf Q_c^{(L)}\mathcal V_L\mathbf Q_c^{(L)}\right)^{-1}
\mathbf Q_c^{(L)}\mathcal V_L\mathbf P_c^{(L)} .
\end{equation}
Under the assumed spectral separation, this contribution is subleading
relative to the projected two-sector structure.

Since the characteristic polynomial depends only on the product $w^{(L)}_{ab}w^{(L)}_{ba}$,
it is convenient to introduce the effective connectivity
\begin{equation}
\omega_{L}=2\sqrt{w^{(L)}_{ab}w^{(L)}_{ba}}.
\end{equation}
In a suitable symmetrized representation, the effective transfer operator
may be written as

\begin{equation}
{\bf V}^{(L)}_{{\rm eff}}=\begin{pmatrix}v^{(L)}_{a} & \omega_{L}/2\\
\omega_{L}/2 & v^{(L)}_{b}
\end{pmatrix}.\label{eq:V_eff}
\end{equation}
The parameter $\omega_{L}$ measures the effective communication between
the competing sectors within the coexistence manifold.

\subsubsection{Spectral splitting and connectivity scale}

Introducing the sector imbalance $\delta_{L}=v^{(L)}_{a}-v^{(L)}_{b}$,
the eigenvalues of Eq.~(\ref{eq:V_eff}) are
\begin{equation}
\lambda^{(L)}_{0,1}=\frac{v^{(L)}_{a}+v^{(L)}_{b}}{2}\pm\frac{1}{2}\sqrt{\delta^{2}_{L}+\omega^{2}_{L}}.\label{eq:Lmb01}
\end{equation}
The corresponding spectral splitting is
\begin{equation}
\Omega_{L}=\lambda^{(L)}_{0}-\lambda^{(L)}_{1}=\sqrt{\delta^{2}_{L}+\omega^{2}_{L}}.\label{eq:Omega}
\end{equation}
Assuming $\bar v_L=\frac{v^{(L)}_a+v^{(L)}_b}{2}$, the dimensionless spectral gap is
\begin{equation}
\Delta_L=\ln\left(\frac{\lambda^{(L)}_0}{\lambda^{(L)}_1}\right)
\approx\frac{\Omega_L}{\bar v_L},
\label{eq:Delta-Omega}
\end{equation}
where the last expression holds for $\Omega_L\ll\bar v_L$. At
spectral balance, $\delta_L=0$, this relation reduces to
$\Delta_L(T_{*,L})\approx\omega_L(T_{*,L})/\bar v_L$.
It is convenient to introduce the dimensionless ratio $r_{L}=\frac{\delta_{L}}{\omega_{L}}$,
which measures the competition between sector imbalance and inter-sector
connectivity.

Thus balance is diagnosed by $\delta_L=0$ (equivalently $r_L=0$
when $\omega_L>0$), whereas $\delta_L\neq0$ signals sector imbalance;
its sign identifies the dominant sector and $|r_L|$ measures the imbalance
relative to the connectivity.

Eq.~(\ref{eq:Omega}) can then be written as
\begin{equation}
\Omega_{L}=\omega_{L}\sqrt{1+r^{2}_{L}}.
\end{equation}

The parameter $r_{L}$ will play a central role in the characterization
of the dominant eigenstates. Large values of $|r_{L}|$ correspond
to imbalance-dominated coexistence, whereas small values of $|r_{L}|$
indicate coexistence controlled primarily by inter-sector connectivity.

\subsubsection{Dominant eigenstate and sector weights}

In the symmetrized representation of Eq.~(\ref{eq:V_eff}), the leading
eigenstates associated with $\lambda^{(L)}_{0,1}$ may be written as
\begin{alignat}{1}
|\psi^{(L)}_{0}\rangle= & \cos\theta_{L}\,|a_{L}\rangle+\sin\theta_{L}\,|b_{L}\rangle,\\
|\psi^{(L)}_{1}\rangle= & -\sin\theta_{L}\,|a_{L}\rangle+\cos\theta_{L}\,|b_{L}\rangle.
\end{alignat}
The mixing angle satisfies $\tan(2\theta_{L})=\frac{\omega_{L}}{\delta_{L}}=\frac{1}{r_{L}}$.
Equivalently, within this symmetrized effective representation,
\begin{alignat}{1}
p_{a}=\cos^{2}\theta_{L}= & \tfrac{1}{2}\left(1+\tfrac{r_{L}}{\sqrt{1+r^{2}_{L}}}\right),\\
p_{b}=\sin^{2}\theta_{L}= & \tfrac{1}{2}\left(1-\tfrac{r_{L}}{\sqrt{1+r^{2}_{L}}}\right).
\end{alignat}

The quantities $p_a$ and $p_b$ are spectral weights in the symmetrized
two-sector basis. For a nonsymmetric transfer operator they should
not be identified directly with probabilities in the original
configuration basis without transforming the corresponding left and
right eigenvectors. For $|r_{L}|\gg1$, one finds
$p_{a}-p_{b}\approx\pm1$, and the dominant
eigenstate is essentially localized within a single sector. In contrast,
when $|r_{L}|\lesssim1$, both sectors acquire comparable spectral
weights in the symmetrized two-sector representation.

\subsubsection{Balanced coexistence and the equal-weight state}\label{subsec:SM1B4}

A special situation occurs when the competing sectors become balanced.
For $\delta_{L}=0$, one has $r_{L}=0$. In this case,
$\theta_{L}=\frac{\pi}{4}$,
and therefore $p_{a}=p_{b}=\frac{1}{2}$. In the symmetrized
two-sector representation, the eigenstates reduce to
\begin{alignat}{1}
|\psi^{(L)}_{0,*}\rangle= & \frac{|a_{L}\rangle+|b_{L}\rangle}{\sqrt{2}},\label{eq:psi_0L}\\
|\psi^{(L)}_{1,*}\rangle= & \frac{|a_{L}\rangle-|b_{L}\rangle}{\sqrt{2}}.
\end{alignat}
The dominant state is thus a unique equal-weight state, with equal
spectral weights in the symmetrized two-sector representation.

This observation provides a natural interpretation of avoided coexistence
phenomena. Rather than being characterized by the crossing of competing
dominant states at finite size, the balance point is marked by the emergence
of a unique balanced dominant state replacing the
degenerate coexistence manifold that appears when $\delta_{L}=0$
and $\omega_{L}=0$.

The splitting $\Omega_L$ controls the finite-size rounding of this
balanced state, while the asymptotic fate of the connectivity decides
whether the low-energy manifold remains irreducible or becomes reducible
in the thermodynamic limit.

\subsection{Pseudo-transitions and avoided coexistence in one dimension}

In strictly one-dimensional systems, the transverse section is trivial.
After the thermodynamic limit $N\to\infty$ is taken, the coexistence
problem is characterized by
\begin{equation}
\delta=v_{a}-v_{b},\qquad\omega=2\sqrt{w_{ab}w_{ba}}.
\end{equation}

The pseudo-transition temperature $T_{*}$ is defined by the balance
condition $\delta(T_{*})=0$. Since the transfer operator remains
irreducible, one has 
\begin{equation}
\Omega(T_{*})=\omega(T_{*})>0.
\end{equation}
Consequently, the coexistence manifold remains spectrally split and
the dominant eigenstate is unique. Therefore, at $T_{*}$ we have
\begin{equation}
|\psi_{0,*}\rangle=\frac{|a\rangle+|b\rangle}{\sqrt{2}},
\end{equation}
corresponding, in the symmetrized representation, to equal sector weights
$p_{a}=p_{b}=\frac{1}{2}$.
Thus, a pseudo-transition is characterized by balanced competing sectors
connected through a finite inter-sector coupling. The coexistence
remains avoided because the connectivity never vanishes.

\subsection{Finite-size coexistence and asymptotic reducibility for \texorpdfstring{$d\geqslant2$}{d >= 2}}

For dimensions $d\geqslant2$ with positive interface tension, the finite-size coexistence manifold
is generated by the sector states $|a_{L}\rangle$ and $|b_{L}\rangle$.
At finite $L$, the coexistence temperature $T_{*,L}$ is determined
by $\delta_{L}(T_{*,L})=0$. 

In the symmetrized representation, the dominant eigenstate then assumes the balanced form
\begin{equation}
|\psi^{(L)}_{0,*}\rangle=\frac{|a_{L}\rangle+|b_{L}\rangle}{\sqrt{2}},
\end{equation}
while the corresponding splitting leads to 
\begin{equation}
\Omega_{L}(T_{*,L})=\omega_{L}(T_{*,L}).
\end{equation}
For coexisting phases with positive interface tension, the Peierls
argument discussed in Sec.~\ref{subsec:sm1A2} gives
\begin{equation}
\omega_{L}=A(L,\beta)e^{-\beta\tau L^{d-1}},\label{eq:omega_L}
\end{equation}
up to subexponential corrections. Therefore,
\begin{equation}
\lim_{L\to\infty}\omega_{L}(T_{*,L})=0,\quad\Rightarrow\quad\lim_{L\to\infty}\Omega_{L}(T_{*,L})=0.
\end{equation}

The finite-size equal-weight state is no longer uniquely selected in
the thermodynamic limit. Instead, the system recovers a degenerate
coexistence manifold associated with two competing pure thermodynamic
phases, represented by the asymptotic states $|a_{\infty}\rangle$
and $|b_{\infty}\rangle$. At the level of the limiting low-energy sector representation, 
the coexistence manifold may therefore be written symbolically as
\begin{equation}
\mathcal{M}^{(\infty)}_{c}={\rm span}\left\{ |a_{\infty}\rangle,|b_{\infty}\rangle\right\}.
\end{equation}
The closing connectivity establishes a reducible thermodynamic
coexistence manifold. A first-order transition additionally requires
a transverse crossing of the restricted free-energy densities,
\begin{equation}
\left.
\frac{\partial}{\partial T}
\left[f_a(T)-f_b(T)\right]
\right|_{T=T_*}\neq0,
\label{eq:transverse-crossing}
\end{equation}
so that the selected branch has a discontinuous entropy. Thus,
first-order coexistence emerges from asymptotic restoration of
reducibility together with a transverse sector crossing.

\subsection{Unified coexistence picture}

The preceding subsections lead to a compact classification. A
one-dimensional pseudo-transition corresponds to
\begin{equation}
\delta(T_*)=0,\qquad \omega(T_*)>0.
\end{equation}
The sectors are balanced, but the finite connectivity keeps the Perron
state unique and the coexistence avoided.

Thermodynamic coexistence with asymptotic restoration of reducibility
requires
\begin{equation}
\lim_{L\to\infty}\delta_L(T_{*,L})=0,\qquad
\lim_{L\to\infty}\omega_L(T_{*,L})=0.
\end{equation}
The spectral splitting then closes and the limiting low-energy manifold
becomes reducible. Finally, the thermodynamic crossing is first order when
condition~(\ref{eq:transverse-crossing}) is satisfied.

Thus balance fixes the coexistence location, connectivity classifies the
finite-size rounding, and the restricted-branch slopes determine whether
the thermodynamic crossing carries a latent heat.

\section{Thermodynamic consequences of avoided coexistence}\label{sec:SM4}

We now investigate the local thermodynamic consequences of avoided
coexistence. Here ``local'' means near the balance point in parameter
and spectral space, not spatially local dynamics. By contrast, global
thermodynamics concerns the dominant eigenvalue of the full transfer
operator over the complete parameter range and its thermodynamic limit.
The relevant low-energy scale is the splitting $\Omega_L$,
which controls the rounding of the finite-size crossover, the enhancement
of response functions, and the growth of the associated spectral length.
The local normal form derived below,
\begin{equation}
\Omega_L(t)=\sqrt{\kappa_L^2t^2+\omega_L^2},\label{eq:Omega(t)}
\end{equation}
implies three useful consequences: the rounded coexistence width
$\Delta T_L\sim\omega_L/|\kappa_L|$, the curvature
$\partial_t^2\Omega_L|_{t=0}=\kappa_L^2/\omega_L$, and the maximal
dominant spectral length $\xi_L^{\rm max}\sim\bar v_L/\omega_L$.

\subsection{Free-energy density near coexistence}

The thermodynamic properties of the coexistence regime are governed
by the dominant eigenvalue of the transfer operator. Within the effective
coexistence manifold constructed in Sec.~\ref{sec:SM1}, the influence
of the competing sectors is encoded in the low-energy splitting $\Omega_{L}$,
which measures the combined effects of sector imbalance and inter-sector
connectivity.

The free-energy density is determined by the dominant eigenvalue,
\begin{equation}
f_{L}=-\frac{1}{\beta L^{d-1}}\ln\lambda^{(L)}_{0}.
\end{equation}
Using the effective coexistence description derived in
Sec.~\ref{subsec:SM1B1},
$\lambda^{(L)}_{0}=\bar{v}_{L}+\frac{1}{2}\Omega_{L}$ with $\bar{v}_{L}=\frac{v^{(L)}_{a}+v^{(L)}_{b}}{2}$,
one obtains
\begin{equation}
f_{L}=-\frac{1}{\beta L^{d-1}}\ln\left(\bar{v}_{L}+\frac{1}{2}\Omega_{L}\right).
\end{equation}

The smooth background contribution is contained in $\bar{v}_{L}$,
whereas the rapid variation associated with coexistence is controlled
by $\Omega_{L}$. Consequently, the thermodynamic anomalies discussed
below originate from the local structure of the coexistence manifold
through its characteristic splitting $\Omega_{L}$. The coexistence
condition is determined by $\delta_{L}=0$, while $\omega_{L}$ controls
the rounding of the finite-size crossover.

\subsection{Local coexistence structure}

Let $t=T-T_{*,L}$ measure the distance from the finite-size coexistence
temperature. For simplicity, temperature is used as the control parameter.
The same local analysis applies to any parameter that drives the coexistence
condition $\delta_{L}=0$.

For finite $L$, the sector weights are analytic functions of the
control parameter. The finite-size coexistence point is defined by
$\delta_{L}(T_{*,L})=0$. We assume that this zero is simple, namely
\begin{equation}
\kappa_{L}=\left.\frac{\partial\delta_{L}}{\partial T}\right|_{T_{*,L}}\neq0.
\end{equation}
Therefore, the local expansion of the imbalance coordinate is
\begin{equation}
\delta_{L}(t)=\kappa_{L}t+\mathcal{O}(t^{2}).
\end{equation}
Over the narrow temperature interval considered below, we assume that
the connectivity varies slowly and approximate
$\omega_L(T)\approx\omega_L(T_{*,L})$. For compactness, the symbol
$\omega_L$ in the following local expressions denotes this value at
the finite-size coexistence temperature.

Eq.~(\ref{eq:Omega(t)}) is the local normal form of the avoided coexistence manifold. 
It shows that the finite-size crossover is controlled by the competition 
between the imbalance scale $|\kappa_L t|$ and the connectivity scale $\omega_L$. 
This competition determines the width of the rounded coexistence core and 
the maximum spectral length reached at balance.

\subsection{Rounded coexistence core}

The rounded coexistence core corresponds to the region in which $\omega_{L}$ varies
slowly with $t$ and
$|t|\lesssim\frac{\omega_{L}}{|\kappa_{L}|}$. The characteristic
width of the rounded coexistence region is therefore $\Delta t_{L}=\Delta T_{L}\sim\frac{\omega_{L}}{|\kappa_{L}|}$.
Differentiating Eq.~(\ref{eq:Omega(t)}),
\begin{equation}
\frac{\partial^{2}\Omega_{L}}{\partial t^{2}}=\frac{\kappa^{2}_{L}\omega^{2}_{L}}{\left(\kappa^{2}_{L}t^{2}+\omega^{2}_{L}\right)^{3/2}},
\end{equation}
which at coexistence yields
\begin{equation}
\left.\frac{\partial^{2}\Omega_{L}}{\partial t^{2}}\right|_{t=0}=\frac{\kappa^{2}_{L}}{\omega_{L}}.
\end{equation}

Consequently, the curvature scale responsible for the enhancement
of response functions is $\kappa_L^2/\omega_L$. The actual response
amplitude may also contain smooth normalization and thermodynamic
factors. The spectral length is determined directly by
\begin{equation}
\xi^{-1}_{L}=\ln\left(\frac{\lambda^{(L)}_{0}}{\lambda^{(L)}_{1}}\right).\label{eq:xi_L}
\end{equation}
We refer to $\xi_L$ as a spectral length because it is defined directly 
from the transfer-matrix spectrum. It coincides with the physical 
correlation length of an observable whenever the subleading mode entering 
Eq. \eqref{eq:xi_L} has a nonzero form factor in the corresponding connected 
correlation function.

Using $\lambda^{(L)}_{0,1}=\bar{v}_{L}\pm\frac{1}{2}\Omega_{L}$,
one obtains
\begin{equation}
\xi^{{\rm max}}_{L}=\Delta^{-1}_{L}
\approx\frac{\bar{v}_{L}}{\omega_{L}}.
\end{equation}
Thus, the maximum spectral length is controlled by the inverse
dimensionless gap, while the response anomalies sharpen as the same
gap closes. Their precise amplitudes additionally depend on
$\kappa_L$ and on smooth thermodynamic prefactors.

At coexistence, the interfacial scaling of the dimensionless gap,
$\Delta_L\sim e^{-\beta\tau L^{d-1}}$, immediately implies
\begin{equation}
\xi^{{\rm max}}_{L}\sim e^{\beta\tau L^{d-1}},
\end{equation}
up to subexponential corrections. This local thermodynamic structure 
is realized explicitly in the decorated bilayer model discussed in Sec. \ref{sec:SM5}.

The separation is controlled when the two leading modes remain
isolated from higher modes and the linear expansion of $\delta_L(T)$ is
accurate. Within the local manifold, sector weights, order parameters, and
sector-resolved correlations distinguish the competing states. Global
behavior is identified from nonanalyticities of the limiting free energy,
discontinuities in its derivatives, and the corresponding long-range order.
We focus on the local neighborhood because it admits a controlled,
model-independent normal form, while the full transfer matrix is retained
in the numerical thermodynamics and determines the model-dependent global
phase diagram.

\section{Microscopic realization and model application}\label{sec:SM5}

\begin{figure}[t]
\includegraphics[width=\columnwidth]{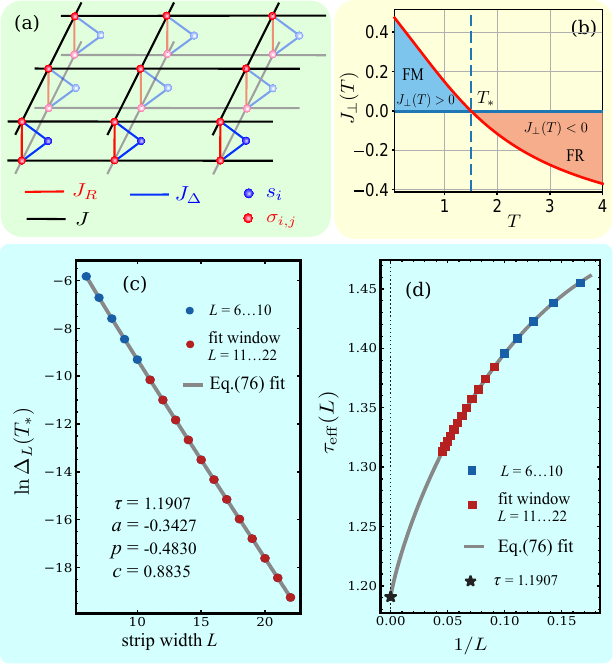}
\caption{Decorated bilayer realization and spectral extraction of the interface tension. (a) Schematic representation of the decorated bilayer Ising strip. (b) Effective interlayer coupling $J_{\perp}(T)$ as a function of temperature, showing the crossover between the FM and FR sectors at $T=T_{*}$. (c) $\ln\Delta_{L}(T_{*})$ versus strip width $L$, with the corrected finite-width fit. (d) Effective interface tension $\tau_{{\rm eff}}(L)=-(T_{*}/L)\ln\Delta_{L}(T_{*})$ versus $1/L$. Parameters: $J=1$, $J_{\Delta}=1.0$, $J_{R}\approx-0.5305$, and $T_{*}=1.5$.}
\label{fig:main-model}
\end{figure}

In this section we present a decorated bilayer Ising model as a microscopic
realization of the coexistence framework developed above. The decoration transformation
generates a temperature-dependent
effective interaction that drives the system across a coexistence
condition between competing ferromagnetic (FM)-like and frustrated
(FR)-like sectors. Figure~\ref{fig:main-model} is the central quantitative
test of the construction: it displays the microscopic model, locates balance
independently through $J_\perp(T_*)=0$, shows the exponential closing of the
coexistence gap, and recovers the exact interface tension from that gap.

\subsection{Decorated bilayer Ising model}\label{subsec:SM3A}

The key ingredient is the decoration transformation~\cite{Dec-trnsf},
which generates a temperature-dependent effective interlayer interaction.
As the temperature varies, this interaction changes sign and drives
the system across a coexistence condition separating competing FM-like
and FR-like sector organizations.

\subsubsection{Decoration transformation}

We consider a decorated bilayer Ising model composed of two coupled
square-lattice layers connected through decorating spins~\cite{chapman-26},
as shown in Fig.~\ref{fig:main-model}a. The Hamiltonian is
\begin{equation}
\mathcal{H}=-J\sum_{\langle ij\rangle,\ell}\sigma_{i,\ell}\sigma_{j,\ell}-J_{R}\sum_{i}\sigma_{i,1}\sigma_{i,2}-J_{\Delta}\sum_{i}s_{i}(\sigma_{i,1}+\sigma_{i,2}),
\end{equation}
where $\sigma_{i,\ell}=\pm1$ ($\ell=\{1,2\}$) denote the nodal spins
on the two layers, while $s_{i}=\pm1$ denotes the decorating spin,
$J$ is the intralayer interaction, $J_{R}$ is the direct interlayer
coupling, and $J_{\Delta}$ couples the decorating spin to the two
nodal spins in the same vertical unit.

The decorating variables can be traced out exactly~\cite{Dec-trnsf},
yielding an effective bilayer model whose interlayer interaction becomes
temperature dependent. For a fixed pair $(\sigma_{i,1},\sigma_{i,2})$,
the local Boltzmann weight becomes
\begin{equation}
\varpi(\sigma_{i,1},\sigma_{i,2})=e^{\beta J_{R}\sigma_{i,1}\sigma_{i,2}}\sum_{s_{i}=\pm1}e^{\beta J_{\Delta}s_{i}(\sigma_{i,1}+\sigma_{i,2})}.
\end{equation}
Performing the sum yields
\begin{equation}
\varpi(\sigma_{i,1},\sigma_{i,2})=2e^{\beta J_{R}\sigma_{i,1}\sigma_{i,2}}\cosh\left[\beta J_{\Delta}(\sigma_{i,1}+\sigma_{i,2})\right].
\end{equation}
Thus we have
\begin{equation}
\varpi(+,+)=\varpi(-,-)=\varpi_{\mathrm{FM}}=2e^{\beta J_{R}}\cosh(2\beta J_{\Delta}),
\end{equation}
and
\begin{equation}
\varpi(+,-)=\varpi(-,+)=\varpi_{\mathrm{FR}}=2e^{-\beta J_{R}}.
\end{equation}
The local weight can therefore be mapped onto an effective Ising interaction,
\begin{equation}
\varpi(\sigma_{i,1},\sigma_{i,2})=\mathcal{A}e^{\beta J_{{\rm \perp}}\sigma_{i,1}\sigma_{i,2}},
\end{equation}
with
\begin{equation}
\mathcal{A}=\sqrt{\varpi_{\rm FM}\varpi_{\rm FR}}
=2\sqrt{\cosh(2\beta J_{\Delta})},
\end{equation}
and
\begin{equation}
J_{\perp}(T)=\frac{T}{2}\ln\frac{\varpi_{{\rm FM}}}{\varpi_{{\rm FR}}}=J_{R}+\frac{T}{2}\ln\left[\cosh\!\left(\frac{2J_{\Delta}}{T}\right)\right].
\end{equation}
The decorated bilayer is therefore mapped exactly onto an effective
bilayer Ising model,
\begin{equation}
\mathcal{H}_{{\rm eff}}=-J\sum_{\langle i,j\rangle}\left(\sigma_{i,1}\sigma_{j,1}+\sigma_{i,2}\sigma_{j,2}\right)-J_{\perp}(T)\sum_{i}\sigma_{i,1}\sigma_{i,2},\label{eq:Heff}
\end{equation}
with a temperature-dependent interlayer coupling.
The local factor $\mathcal A(T)$ contributes an additive term
$-T\ln\mathcal A(T)$ per vertical unit to the free energy. It does
not affect nodal-spin correlations or the balance condition, but it
must be retained when calculating entropy and specific heat; this
contribution is included in the numerical thermodynamic curves below.

The sign of $J_{\perp}(T)$ determines the dominant low-energy organization
of the bilayer. For $J_{\perp}(T)>0$, parallel interlayer alignment
is favored, defining an FM-like sector. Conversely, $J_{\perp}(T)<0$
favors antiparallel interlayer alignment and defines an FR-like sector.
The balance point between these competing sector organizations is
determined by $J_{\perp}(T_{*})=0$.

This condition identifies the temperature at which the FM-like and
FR-like sectors become equally favored. In the language of
Sec.~\ref{sec:SM1},
it provides the microscopic realization of sector balance. The two
couplings appearing in Eq.~(\ref{eq:Heff}) play distinct roles: the
effective interlayer coupling $J_{\perp}(T)$ controls the competition
between sectors, whereas the intralayer coupling $J$ controls their
spatial coherence and the energetic cost of interfaces separating
competing sector organizations. As shown below, these two ingredients
give rise naturally to the coexistence coordinates $\delta_{L}$ and
$\omega_{L}$ introduced in Sec.~\ref{sec:SM1}.

\subsubsection{Restricted sectors and spectral balance}

Near the balance point $T_{*}$, the low-energy spectrum is dominated
by two competing sector organizations, namely the FM-like and FR-like
sectors identified above. The full strip transfer operator $V_{L}$
acts on a configuration space whose dimension grows exponentially
with the strip width $L$. However, close to sector balance, the dominant
spectral structure is controlled by the leading restricted states
associated with these two competing organizations.

Let $|\mathrm{FM}_{L}\rangle$ and $|\mathrm{FR}_{L}\rangle$ denote
the dominant right eigenvectors of the transfer operators restricted
to the FM-like and FR-like sectors, respectively. The relevant low-energy
structure is then captured by the two-dimensional coexistence manifold
\begin{equation}
\mathcal{M}^{(L)}_{c}={\rm span}\Big\{|\mathrm{FM}_{L}\rangle,|\mathrm{FR}_{L}\rangle\Big\}.
\end{equation}

Under the spectral-separation assumptions stated in
Sec.~\ref{subsec:SM1B1}, projection onto this manifold yields the
effective two-sector description
\begin{equation}
V^{(L)}_{{\rm eff}}=\begin{pmatrix}v^{(L)}_{{\rm FM}} & \omega_{L}/2\\
\omega_{L}/2 & v^{(L)}_{{\rm FR}}
\end{pmatrix},
\end{equation}
where $v^{(L)}_{{\rm FM}}$ and $v^{(L)}_{{\rm FR}}$ are the dominant
eigenvalues of the transfer operators restricted to the FM-like and
FR-like sectors, while the off-diagonal elements describe residual
communication between the two sector organizations. This reduced operator
is not used to replace the full transfer matrix in the numerical calculations.
Rather, it provides the effective low-energy description that organizes
the dominant spectrum near coexistence.

For strip geometry, $d=2$, so that the transverse volume entering
the general framework of Sec.~\ref{sec:SM1} reduces to $L^{d-1}=L$. At
leading exponential order, the restricted-sector eigenvalues may be
expressed in terms of the corresponding restricted free-energy densities
as
\begin{equation}
v^{(L)}_{{\rm FM}}\approx e^{-\beta Lf_{{\rm FM}}(T)},\qquad v^{(L)}_{{\rm FR}}\approx e^{-\beta Lf_{{\rm FR}}(T)}.
\end{equation}

The following expressions are the leading ordered-sector
approximations to the restricted free-energy densities. For parallel
interlayer alignment, the local vertical weight is
$\varpi_{{\rm FM}}=2e^{\beta J_{R}}\cosh(2\beta J_{\Delta})$,
whereas for antiparallel interlayer alignment,
$\varpi_{{\rm FR}}=2e^{-\beta J_{R}}$,
because the decorating spin remains free. Approximating each ordered
layer by its dominant aligned configuration gives
\begin{equation}
f_{{\rm FM}}(T)\approx-4J-J_{R}-T\ln\!\left[2\cosh\!\left(\frac{2J_{\Delta}}{T}\right)\right],
\end{equation}
and
\begin{equation}
f_{{\rm FR}}(T)\approx-4J+J_{R}-T\ln2.
\end{equation}

The term $-T\ln2$ reflects the residual entropy associated with the
free decorating spin in the FR-like sector. At this leading
ordered-sector level, the difference between the two restricted
free-energy densities is
\begin{equation}
f_{{\rm FM}}(T)-f_{{\rm FR}}(T)=-2J_{R}-T\ln\left[\cosh\!\left(\frac{2J_{\Delta}}{T}\right)\right]=-2J_{\perp}(T).
\end{equation}
Introducing the average restricted free-energy density
\begin{equation}
\bar{f}(T)=\frac{f_{{\rm FM}}(T)+f_{{\rm FR}}(T)}{2}=-4J-\frac{T}{2}\ln\!\left[4\cosh\!\left(\frac{2J_{\Delta}}{T}\right)\right],
\end{equation}
the restricted-sector eigenvalues become
\begin{equation}
v^{(L)}_{{\rm FM}}\approx e^{-\beta L\bar{f}(T)}e^{\beta LJ_{\perp}(T)},\qquad v^{(L)}_{{\rm FR}}\approx e^{-\beta L\bar{f}(T)}e^{-\beta LJ_{\perp}(T)}.
\end{equation}

The spectral imbalance coordinate introduced in Sec.~\ref{sec:SM1} is
therefore realized microscopically as
\begin{equation}
\delta_{L}=v^{(L)}_{{\rm FM}}-v^{(L)}_{{\rm FR}}\approx2e^{-\beta L\bar{f}(T)}\sinh\!\bigl[\beta LJ_{\perp}(T)\bigr].
\end{equation}
Consequently,
\begin{equation}
\delta_{L}=0\quad\Rightarrow\quad J_{\perp}(T_{*})=0,
\end{equation}
showing that the temperature-driven sign change of the effective
interlayer coupling is realized spectrally as the vanishing of the
imbalance coordinate. More generally, the effective bilayer is
invariant under flipping all spins in one layer together with
$J_\perp\to-J_\perp$; hence its two ordered restricted branches are
exactly balanced at $J_\perp=0$, independently of the approximation
used above.

For the parameters used below, $T_*=1.5$ lies below the ordering
temperature of each decoupled square-lattice layer,
$T_c=2J/\ln(1+\sqrt{2})\approx2.269$. The crossing is transverse
because $J'_\perp(T_*)\neq0$. The thermodynamic transition occurs at
$T=T_*=1.5<T_c$ and is determined by $J_\perp(T_*)=0$. This zero is shown in
Fig.~\ref{fig:main-model}b, and the vertical dashed line in
Fig.~\ref{fig:supmat-1} marks the same temperature. At finite width this
point appears as an analytic rounded crossover; the first-order singularity appears only as
$L\to\infty$, when the connectivity vanishes. In the thermodynamic ordered phase, the
one-sided restricted entropies differ by
\begin{equation}
s_{\rm FM}-s_{\rm FR}
=2m^2(T_*)J'_\perp(T_*),
\label{eq:entropy-jump}
\end{equation}
where $m(T_*)$ is the spontaneous magnetization of one decoupled
layer. Thus the closing connectivity is accompanied by a nonzero
entropy jump and latent heat. The fully aligned approximation above
corresponds to setting $m=1$.

The remaining ingredient of the coexistence manifold is the inter-sector
connectivity $\omega_{L}$, which is not determined by the restricted
free energies alone. Instead, it originates from the residual communication
between the FM-like and FR-like sector organizations and is discussed
in the next subsection.

\subsubsection{Inter-sector connectivity}\label{subsec:SM3A3}

The previous subsection showed that the balance condition $J_{\perp}(T_{*})=0$
corresponds to the vanishing of the imbalance coordinate $\delta_{L}$.
The second coexistence coordinate $\omega_{L}$ originates from residual
communication between the competing FM-like and FR-like sector organizations.

Near sector balance, communication between the two sectors is mediated
by configurations containing interfaces separating FM-like and FR-like
regions. In strip geometry, the dominant contribution arises from
a domain wall spanning the transverse direction of the system. If
$\tau$ denotes the corresponding effective interfacial free-energy
cost per unit transverse length, the interface free energy scales
as $F_{{\rm int}}\approx\tau L$, giving a statistical weight $e^{-\beta F_{{\rm int}}}=e^{-\beta\tau L}$.

Since the off-diagonal amplitudes $w^{(L)}_{{\rm FM,FR}}$ and $w^{(L)}_{{\rm FR,FM}}$
of the effective operator $V^{(L)}_{{\rm eff}}$ are generated by
such interface-mediated processes, one expects
\begin{equation}
\omega_{L}\sim e^{-\beta\tau L}.
\end{equation}

Thus, the two microscopic couplings generate the two coordinates of
the coexistence manifold: the effective interlayer coupling $J_{\perp}(T)$
controls the imbalance coordinate $\delta_{L}$, whereas the intralayer
coupling $J$ controls the connectivity coordinate $\omega_{L}$ through
the interfacial free-energy cost. As shown below, the finite-size
scaling of the dominant spectral gap provides a direct numerical probe
of this connectivity scale.

\subsection{Numerical realization of the coexistence manifold}
\label{subsec:SM3B}

The numerical results presented in this section provide direct evidence
for the spectral mechanism proposed in the article. In particular,
they demonstrate exponentially suppressed inter-sector connectivity,
weakly connected low-energy manifolds, avoided-crossing scaling,
and progressive restoration of reducibility with increasing
strip width.

The operational point is that the gap is evaluated at the independently
known balance condition $J_\perp(T_*)=0$. It therefore isolates
connectivity rather than merely identifying a small gap. Together,
the results show that the coexistence regime is governed by weakly
connected low-energy sectors whose communication decreases
exponentially with system size.

\subsubsection{Thermodynamic manifestation of sector balance}

\begin{figure}[t]

\includegraphics[scale=0.5]{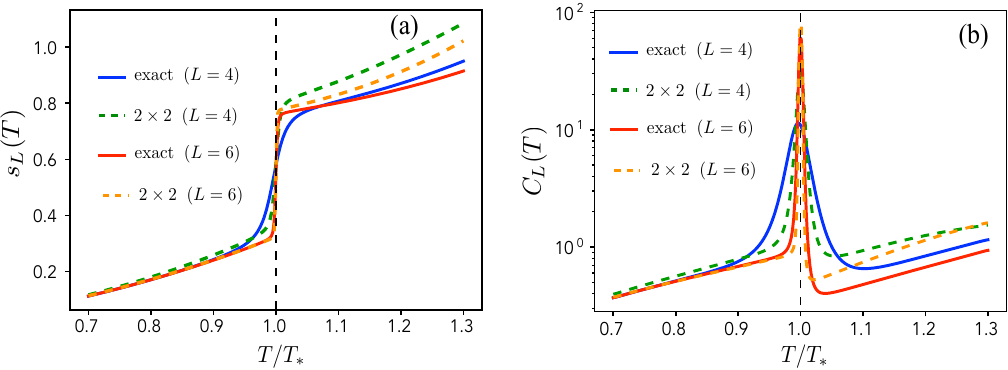}
\caption{Thermodynamic signatures of finite-size coexistence. (a) Entropy
per site $s_{L}(T)$ as a function of $T/T_{*}$ for strip widths $L=4$
and $L=6$, comparing the full transfer-matrix result with the effective
two-sector ($2\times2$) description. The vertical dashed line indicates
the balance temperature $T=T_{*}$. (b) Specific heat per site $C_{L}(T)$
on a logarithmic scale for the same strip widths. The sharpening of
the entropy crossover and of the specific-heat anomaly with increasing
$L$ is consistent with the narrowing of the finite-size coexistence
window. Parameters: $J=1$, $J_{\Delta}=1.0$, $J_{R}\approx-0.5305$,
and $T_{*}=1.5$.}\label{fig:supmat-1}
\end{figure}

Fig.~\ref{fig:supmat-1}a shows the entropy per
site for two strip widths $L$. The entropy exhibits
a rapid crossover near the balance temperature $T_{*}$, where the
competing FM-like and FR-like sector organizations become thermodynamically
equivalent. For every finite strip width, the entropy remains continuous
and fully analytic. However, the crossover becomes progressively sharper
as $L$ increases. As discussed in Sec.~\ref{sec:SM1} and \ref{sec:SM4}, this behavior is
expected from the exponential suppression of the inter-sector connectivity:
as the connectivity decreases, the coexistence window narrows and
the thermodynamic crossover becomes increasingly abrupt.

At low temperatures, the thermodynamics is governed by the FM-like
sector, which minimizes the restricted free energy when $J_{\perp}(T)>0$.
As the temperature increases, the entropic contribution associated
with the free decorating spins progressively lowers the FR-like free
energy. The balance condition $J_{\perp}(T_{*})=0$ marks the crossover
between the two competing sector organizations. The finite-size evolution
visible in Fig.~\ref{fig:supmat-1}a therefore provides
a direct thermodynamic manifestation of sector balance: the crossover
sharpens systematically with increasing strip width, while remaining
fully analytic for every finite strip width.

Fig.~\ref{fig:supmat-1}b shows the corresponding
specific heat for the same strip widths. As $L$ increases, the
specific-heat peak becomes simultaneously
higher and narrower around $T_{*}$, reflecting the progressive suppression
of the finite-size connectivity between the competing sectors. Nevertheless,
the peak remains finite for every finite strip width, consistent with
the analyticity of the free energy and with the avoided-coexistence
structure discussed in Secs.~\ref{sec:SM1} and \ref{sec:SM4}.
The agreement between the full transfer-matrix calculation and the
effective two-sector description shows that the same low-energy
manifold controlling the spectral gap also captures the thermodynamic
sharpening near balance. Away from the coexistence window, the reduced
description is not expected to remain exact because higher transfer-matrix
modes are no longer negligible.

Consistently with the local rounding analysis of
Sec.~\ref{sec:SM4},
the growth and narrowing of the specific-heat peak are controlled by
the closing dimensionless gap $\Delta_L$. The precise peak amplitude
also depends on the temperature derivative of the sector imbalance
and on smooth thermodynamic prefactors. The narrowing of the peak
demonstrates that the coexistence regime becomes confined to an
increasingly small temperature window as the system approaches
asymptotic reducibility.

\subsubsection{Numerical probe of inter-sector connectivity}

Fig.~\ref{fig:main-model}c shows $\ln\Delta_{L}(T_{*})$, where
$\Delta_{L}=\ln(\lambda_{0}/\lambda_{1})$
is the dimensionless spectral gap of the full strip transfer matrix.
At coexistence, Eq.~(\ref{eq:Delta-Omega}) gives
$\Delta_{L}(T_{*})\approx\omega_{L}(T_{*})/\bar{v}_{L}$.
Thus, $\Delta_{L}(T_{*})$ probes the residual inter-sector connectivity
normalized by the dominant sector weight.

Since this residual connectivity is generated by interface-mediated
communication between the competing FM and FR sector organizations,
the central quantitative prediction is
\begin{equation}
\Delta_{L}(T_{*})\sim A(L,T_{*})e^{-\beta_{*}\tau L},
\end{equation}
where $A(L,T_{*})$ absorbs the connectivity prefactor and the normalization
by $\bar{v}_{L}$, and contains only subexponential corrections. Equivalently,
\begin{equation}
\ln\Delta_{L}(T_{*})=-\beta_{*}\tau L+\ln A(L,T_{*}).
\end{equation}

The deviations observed at finite $L$ contain preasymptotic interface
and prefactor effects. Such corrections are a central issue in
conventional finite-size-scaling analyses of first-order transitions
and interface-tension extraction~\cite{MataragkasFSS25,Mataragkas3D26};
here they enter only through the prefactor multiplying the spectral
connectivity. We define
\begin{equation}
\tau_{{\rm eff}}(L)=-\frac{T_{*}}{L}\ln\Delta_{L}(T_{*}).
\end{equation}
If the prefactor were $L$ independent, a linear extrapolation would
give
\begin{equation}
\tau_{{\rm eff}}(L)=\tau_{\infty}+\frac{c_{2}}{L}+o(L^{-1}),
\end{equation}
and the window $L=11,\ldots,22$ yields the finite-window estimate
$\tau_{\rm lin}=1.241$. This number should not be interpreted as a
statistically exact thermodynamic extrapolation, because
$A(L,T_*)$ is not constant.

At $J_\perp(T_*)=0$, the effective model consists of two decoupled
square-lattice Ising layers. The exact axial interface tension of one
layer is
\begin{equation}
\tau_{\rm Ising}(T_*)
=2J-T_*\ln\!\left[\coth\!\left(\frac{J}{T_*}\right)\right]
=1.19009\ldots.
\label{eq:tau-exact}
\end{equation}
To resolve the finite-width bias, we fit the same data to the
minimal subexponential expansion containing the leading algebraic
factor and the first inverse-width correction. Equivalently, we write
the prefactor as
\begin{equation}
A(L,T_*)=A_0L^p\exp(c/L),
\end{equation}
so that
\begin{equation}
\ln\Delta_L
=a-\beta_*\tau L+p\ln L+\frac{c}{L}.
\label{eq:gap-corrected-fit}
\end{equation}
The value of $\tau$ is not imposed in this fit. For
$L=11,\ldots,22$, it gives $\tau=1.1907$; shifting the lower cutoff
to $L=14$ gives $\tau=1.1904$. The value $\tau=1.1907$ agrees with
$\tau_{\rm Ising}=1.19009\ldots$ in Eq.~(\ref{eq:tau-exact}) to
better than $7\times10^{-4}$. Thus the
data demonstrate exponential suppression of connectivity and recover
the exact interface tension after finite-width corrections are
resolved. The nontrivial result is not the known Ising value itself,
but its extraction from the coexistence spectral gap without
constructing or imposing an interface. The linear value $1.241$
should therefore be read only as an effective finite-window slope
obtained when the subexponential prefactor is held constant.

Table~\ref{tab:fit-stability} tests whether this agreement is produced
by an overly flexible fit. All entries use the same data up to
$L_{\max}=22$ and unweighted least squares. A pure exponential shows
a visible finite-window bias. Including the algebraic prefactor reduces
the residuals substantially, while the inverse-width correction is
needed to recover the exact tension. Within the complete form, changing
$L_{\min}$ from 11 to 14 changes $\tau$ by only $2.6\times10^{-4}$,
and the residual decreases monotonically as preasymptotic widths are
removed.

\begin{table}[t]
\caption{Stability of the interface-tension extraction. Here
$R=[N^{-1}\sum_L r_L^2]^{1/2}$ is the root-mean-square residual of
$\ln\Delta_L$, and $\tau_{\rm Ising}=1.19009\ldots$.}
\label{tab:fit-stability}

\begin{tabular}{lccc}
\hline
fit form & $L_{\min}$ & $\tau$ & $R$\\
\hline
$a-\beta_*\tau L$ & 11 & 1.2411 & $1.24\times10^{-2}$\\
$a-\beta_*\tau L+p\ln L$ & 11 & 1.1853 & $1.47\times10^{-4}$\\
$a-\beta_*\tau L+p\ln L+c/L$ & 11 & 1.1907 & $3.27\times10^{-6}$\\
same & 12 & 1.1906 & $1.59\times10^{-6}$\\
same & 13 & 1.1905 & $8.02\times10^{-7}$\\
same & 14 & 1.1904 & $5.10\times10^{-7}$\\
\hline
\end{tabular}

\end{table}

\subsubsection{Low-lying spectrum and block spectral length}

\begin{figure}[t]
\includegraphics[scale=0.5]{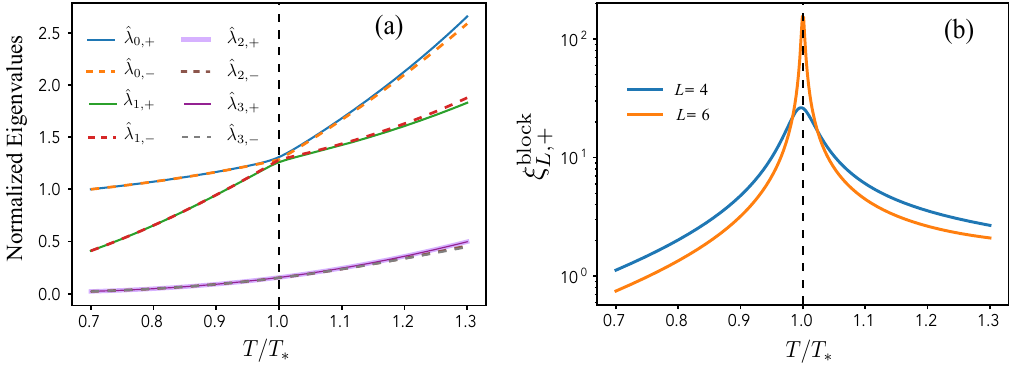}
\caption{Low-lying parity towers and block spectral length. 
(a) Leading normalized transfer-matrix
eigenvalues $\hat{\lambda}_{n,+}$ (solid lines) and
$\hat{\lambda}_{n,-}$ (dashed lines) for $L=4$ as functions of
$T/T_{*}$, where $\pm$ denotes the global spin-flip parity. The
low-energy spectrum organizes into nearly degenerate parity pairs.
(b) Block spectral length $\xi^{\mathrm{block}}_{L,+}$, defined by
Eq.~(\ref{eq:xi-block}), in the even global spin-flip parity block
for $L=4$ and $L=6$. The vertical axis is logarithmic. Parameters:
$J=1$, $J_{\Delta}=1.0$, $J_{R}\approx-0.5305$, and
$T_{*}=1.5$. Panel (a) reveals the symmetry-resolved organization
of the low-energy manifold, whereas panel (b) shows the growth of
the longest coexistence-related scale after the global spin-flip
sector has been fixed.}\label{fig:supmat-2}
\end{figure}

The coexistence gap analyzed above isolates the communication between
the FM-like and FR-like organizations. A complementary question is how
this avoided coexistence is embedded in the full low-energy spectrum.
Figure~\ref{fig:supmat-2} shows that the spectrum develops
symmetry-resolved parity towers and, within a fixed parity block, a
rapidly growing spectral length near balance.
In Fig.~\ref{fig:supmat-2}a,
the leading normalized eigenvalues, $\hat{\lambda}_{j}=\lambda_{j}e^{\beta E_{0}}$,
organize into nearly degenerate pairs,
\begin{equation}
(\hat{\lambda}_{0,+},\hat{\lambda}_{0,-}),\quad
(\hat{\lambda}_{1,+},\hat{\lambda}_{1,-}),\quad
(\hat{\lambda}_{2,+},\hat{\lambda}_{2,-}),\ldots,
\end{equation}
where $\pm$ denotes the global spin-flip parity.
This doublet structure reflects the internal organization of the
ordered spectrum into symmetry-related components.

The block length in Fig.~\ref{fig:supmat-2}b is distinct from the
dominant spectral length $\xi_L$ in Eq.~(\ref{eq:xi_L}). The latter
is associated with the gap of the full transfer operator, whereas
$\xi^{\rm block}_{L,+}$ removes the near degeneracy associated with
global spin inversion and measures the remaining scale inside the
even-parity block.
We define
\begin{equation}
\xi^{\mathrm{block}}_{L,+}(T)
=\left[
\ln\left(\frac{\lambda^{(L)}_{0,+}(T)}
{\lambda^{(L)}_{1,+}(T)}\right)
\right]^{\!-1},
\label{eq:xi-block}
\end{equation}
where $\lambda^{(L)}_{0,+}$ and $\lambda^{(L)}_{1,+}$ are the two
largest eigenvalues in that block. The quantity
$\xi^{\mathrm{block}}_{L,+}$ measures the longest spectral scale
that remains after fixing the global spin-flip parity. Its enhancement
near $T/T_{*}=1$, and its rapid growth from $L=4$ to $L=6$, show the
finite-size sharpening of the coexistence regime.

The block spectral length $\xi^{\rm block}_{L,+}$ becomes a correlation 
length for observables within the even-parity sector that couple to the 
mode $\lambda^{(L)}_{1,+}$. The interlayer observable
$\sigma_{i,1}\sigma_{i,2}$, being even under global spin inversion
and directly distinguishing FM-like from FR-like organization, is the
natural observable expected to couple to this mode.

Thus, Fig.~\ref{fig:supmat-2} provides complementary evidence for
the spectral organization expected from weak irreducibility. Panel
(a) displays the near-degenerate parity towers, while panel (b)
shows directly how the longest scale inside a fixed symmetry block
grows with strip width. The residual connectivity between the competing
FM and FR organizations is quantified separately by the dominant
coexistence gap $\Delta_{L}(T_{*})$ analyzed in Fig.~\ref{fig:main-model}c.
Fig.~\ref{fig:supmat-2} should therefore be read as
evidence for the broader ordered-sector doublet structure, whereas
Fig.~\ref{fig:main-model}c isolates the specific connectivity scale
controlling the avoided coexistence between the FM and FR sectors.

\section{Discussion and conclusions}

The spectral framework developed above gives a unified operator-level
interpretation of pseudo-transitions, finite-size avoided coexistence,
and thermodynamic first-order coexistence. It is consistent with
restricted-ensemble, finite-size, and partition-function-zero approaches
\cite{Pirogov76,Borgs,Binder87,Lee,Pirogov75,bricmont,Friedli,Binder84,biskup,Cannon24,YangLee,LeeYang,Moueddene25},
but focuses on the low-energy transfer operator itself: after sector
balance has been located, the remaining spectral gap measures residual
communication between the sectors.

This distinction is especially useful for comparing pseudo-transitions with
true thermodynamic coexistence. In a one-dimensional finite-state system,
sector balance can occur at finite temperature, but finite connectivity keeps
the Perron state unique and the coexistence avoided. In contrast, for
short-range systems with positive interface tension, the interface cost
suppresses the connectivity exponentially with transverse size. The gap then
closes, the low-energy sector becomes reducible, and distinct thermodynamic
phases emerge. A first-order transition additionally requires a transverse
crossing of the restricted free-energy branches.

The decorated bilayer realizes these ingredients independently. The
temperature-dependent effective interlayer coupling cancels the imbalance at
$T_*$, while interfacial suppression removes the residual connectivity as the
strip width grows. Quantitatively, the coexistence gap at independently
located balance recovers the exact Ising interface tension, with
$\tau=1.1907$ compared with
$\tau_{\rm Ising}=1.19009\ldots$. The same two-sector manifold also
reproduces the entropy and specific-heat sharpening near balance, while the
full low-energy spectrum organizes into parity-resolved near doublets and
shows a growing block spectral length $\xi^{\rm block}_{L,+}$.

Thus the central criterion can be stated compactly. Balance locates the
coexistence point, connectivity classifies it, and the restricted-branch
slopes determine whether the thermodynamic crossing is first order. The
recovery of the exact interface tension, the agreement between the full and
reduced thermodynamics, and the symmetry-resolved growth of the low-energy
spectral length show that weak irreducibility provides both a classification
principle and a quantitative probe of finite-size coexistence.
\section*{Acknowledgements}
This work was partially supported by the Brazilian agencies CNPq, CAPES, and FAPEMIG.

\appendix

\section{Balanced-branch entropy density and coexistence states}\label{subsec:SM1D}

The main text used spectral balance to locate coexistence and the
connectivity to classify its finite-size rounding. This appendix records
two consequences of the same balanced branch: the residual entropy density
obtained from the eigenvalue and the finite-temperature balanced state
obtained from the eigenvector.

\subsection{Residual balanced-branch entropy density and order of limits}\label{subsec:SM1D1}

The entropy density associated with the dominant eigenvalue is
\begin{equation}
s^{(L)}
=\frac{1}{L^{d-1}}
\left[
\ln\lambda^{(L)}_{0}
-\beta\frac{d}{d\beta}\ln\lambda^{(L)}_{0}
\right].
\label{eq:s-def}
\end{equation}
The temperature dependence of $\lambda^{(L)}_{0}$ enters through the
sector weights $v^{(L)}_{a}$ and $v^{(L)}_{b}$ and the connectivity
$\omega_{L}$; hence
\begin{equation}
\frac{d\lambda^{(L)}_{0}}{d\beta}=\frac{\partial\lambda^{(L)}_{0}}{\partial v^{(L)}_{a}}\frac{d v^{(L)}_{a}}{d\beta}+\frac{\partial\lambda^{(L)}_{0}}{\partial v^{(L)}_{b}}\frac{d v^{(L)}_{b}}{d\beta}+\frac{\partial\lambda^{(L)}_{0}}{\partial\omega_{L}}\frac{d\omega_{L}}{d\beta}.\label{eq:diff-L0}
\end{equation}
Using Eq.~(\ref{eq:Lmb01}), one finds
\begin{equation}
\frac{\partial\lambda^{(L)}_{0}}{\partial v^{(L)}_{a}}=\frac{1}{2}+\frac{\delta_{L}}{2\Omega_{L}},\qquad
\frac{\partial\lambda^{(L)}_{0}}{\partial v^{(L)}_{b}}=\frac{1}{2}-\frac{\delta_{L}}{2\Omega_{L}},\qquad
\frac{\partial\lambda^{(L)}_{0}}{\partial\omega_{L}}=\frac{\omega_{L}}{2\Omega_{L}}.
\end{equation}
At spectral balance, $\delta_{L}=0$, the dominant eigenvalue therefore
responds symmetrically to the two sector weights:
\begin{equation}
\frac{\partial\lambda^{(L)}_{0}}{\partial v^{(L)}_{a}}\biggr|_{\delta_{L}=0}=\frac{\partial\lambda^{(L)}_{0}}{\partial v^{(L)}_{b}}\biggr|_{\delta_{L}=0}=\frac{1}{2}.\label{eq:sym-1/2}
\end{equation}

We now consider a low-temperature regime in which the restricted
eigenvalues have the asymptotic form
\begin{equation}
v^{(L)}_{\alpha}
=\exp\left\{
L^{d-1}\left[s^{(L)}_{\alpha}
-\beta e^{(L)}_{\alpha}\right]
\right\},
\qquad \alpha=a,b,
\label{eq:v_ab-s_ab}
\end{equation}
where $s^{(L)}_{\alpha}$ and $e^{(L)}_{\alpha}$ are the leading
temperature-independent residual entropy and energy densities. Thus,
$d v^{(L)}_{\alpha}/d\beta
=-L^{d-1}e^{(L)}_{\alpha}v^{(L)}_{\alpha}$.
Along the balanced branch,
$v^{(L)}_{a}=v^{(L)}_{b}\equiv v^{(L)}_{c}$. We take the residual
weak-connectivity limit along this branch,
\begin{equation}
\frac{\omega_{L}}{v^{(L)}_{c}}\longrightarrow0,\qquad
\frac{\beta}{L^{d-1}v^{(L)}_{c}}
\frac{d\omega_{L}}{d\beta}\longrightarrow0.
\label{eq:omega-entropy-limit}
\end{equation}
The second condition ensures that the connectivity makes no residual
contribution to the entropy density. Equations~(\ref{eq:s-def}),
(\ref{eq:diff-L0}), and (\ref{eq:sym-1/2}) then give
\begin{equation}
s^{(L)}_{c}
=\frac{\ln v^{(L)}_{c}}{L^{d-1}}
+\frac{\beta}{2}\left(e^{(L)}_{a}
+e^{(L)}_{b}\right).
\label{eq:s-c-intermediate}
\end{equation}
Meanwhile, Eq.~(\ref{eq:v_ab-s_ab}) and spectral balance imply
\begin{equation}
\frac{\ln v^{(L)}_{c}}{L^{d-1}}
=\frac{s^{(L)}_{a}+s^{(L)}_{b}}{2}
-\frac{\beta}{2}\left(e^{(L)}_{a}
+e^{(L)}_{b}\right).
\end{equation}
Substitution into Eq.~(\ref{eq:s-c-intermediate}) cancels the energy
terms, yielding
\begin{equation}
s^{(L)}_{c}
=\frac{s^{(L)}_{a}+s^{(L)}_{b}}{2}.
\label{eq:s_cL}
\end{equation}
Its affine form is closely related to the thermodynamic structure of
first-order coexistence. In a reducible infinite-volume coexistence
manifold, the convex Gibbs state
$\mu_x=x\mu_a+(1-x)\mu_b$ has entropy density
\begin{equation}
s_{\rm mix}(x)=xs_{a}+(1-x)s_{b},
\qquad 0\leqslant x\leqslant1,
\label{eq:coexistence-entropy-mixture}
\end{equation}
and therefore
\begin{equation}
\min\left\{s_{a},s_{b}\right\}
\leqslant s_{\rm mix}(x)
\leqslant\max\left\{s_{a},s_{b}\right\}.
\label{eq:coexistence-entropy-bounds}
\end{equation}
Here the affine relation applies to the thermodynamic entropy density.
The same bulk relation describes phase-separated states with volume
fractions $x$ and $1-x$. A finite-volume entropy may additionally
contain an $O(1)$ Gibbs-Shannon mixing term and subextensive
interfacial corrections.

Equation~(\ref{eq:s_cL}) has the symmetric form associated with
$x=1/2$, but this correspondence is structural and limiting. At finite
$L$ and $\omega_{L}>0$, the dominant transfer-operator state is unique
and hybridized, not a macroscopic spatial mixture of two phases. The
phase-mixture interpretation applies only after the appropriate
reducible thermodynamic limit has restored the coexistence manifold.
The arithmetic-mean identity, previously obtained for $d=1$ in
Ref.~\cite{phya-o25}, instead follows from a definite order of limits:
one first imposes spectral balance at finite $\omega_{L}>0$ and then
takes the residual weak-connectivity limit in
Eq.~(\ref{eq:omega-entropy-limit}). Hence $s^{(L)}_{c}$ is a
balanced-branch entropy density, not the entropy density of a third
thermodynamic phase.

The opposite prescription is to impose strict reducibility first\cite{ph-bd}.
Assuming that $a$ and $b$ are the only dominant sectors, one then has
\begin{equation}
\lambda^{(L)}_{0}=\max\left\{v^{(L)}_{a},v^{(L)}_{b}\right\}.
\label{eq:lambda-reducible-selection}
\end{equation}
Away from the exact crossing, the larger restricted eigenvalue selects
one block. At a finite-temperature crossing with different slopes, the equilibrium 
free energy is continuous but nondifferentiable, and its two one-sided 
derivatives give the pure-sector entropy densities. Exactly at 
thermodynamic coexistence, the
convex Gibbs states in Eq.~(\ref{eq:coexistence-entropy-mixture}) remain
available. In the residual $T\to0^{+}$ selection problem, if the
competing sectors have the same leading energy, the larger residual
entropy density selects the dominant block:
\begin{equation}
s^{(L)}_{\rm sel}
=\max\left\{s^{(L)}_{a},s^{(L)}_{b}\right\}.
\label{eq:s-reducible-selection}
\end{equation}
Thus, the two prescriptions do not commute. $s^{(L)}_{c}$ is obtained
by continuing along the balanced weakly irreducible branch before the
connectivity vanishes, whereas $s^{(L)}_{\rm sel}$ results from
pure-block selection after reducibility has been imposed. Their
relation is
\begin{equation}
s^{(L)}_{c}
=\frac{s^{(L)}_{a}+s^{(L)}_{b}}{2}
\leqslant s^{(L)}_{\rm sel}.
\label{eq:s-balanced-bound}
\end{equation}
The balanced branch therefore has no residual-entropy excess over the
adjacent sectors. It describes avoided coexistence between them, not an
additional entropy-dominated sector. An intermediate manifold whose
residual entropy density exceeded both $s^{(L)}_{a}$ and
$s^{(L)}_{b}$ would
instead constitute an additional thermodynamic sector and would fall
outside the two-sector mechanism.

Finally, Eq.~(\ref{eq:s-def}) also defines a balanced-branch entropy
density at a generic finite balance temperature. The arithmetic-mean
identity, however, is specific to the residual regime above, because
at finite temperature the restricted-sector entropy densities and the
connectivity contribution retain their full temperature dependence.

\subsection{Finite-temperature balanced state and residual limit}

The eigenvector counterpart of the balanced-branch entropy is the 
finite-temperature balanced state already obtained in Sec. \ref{subsec:SM1B4}. 
At spectral balance, $\delta_L=0$, and for finite connectivity 
$\omega_L>0$, the mixing angle becomes $\theta_L=\pi/4$. The 
dominant eigenstate therefore assumes, in the symmetrized representation, the balanced form Eq.~(\ref{eq:psi_0L}).
This state has equal spectral weights in the symmetrized two-sector
representation and constitutes the eigenstate counterpart of the
symmetric spectral response discussed in \ref{subsec:SM1D1}.

Along the balanced branch, the dominant state has the equal-weight
form in Eq.~(\ref{eq:psi_0L}) whenever the connectivity is
finite. In a residual limit $T\rightarrow0^{+}$,
the connectivity $\omega_{L}$ may vanish and sector degeneracy is
progressively restored. The coexistence state $|\psi^{(L)}_{0,*}\rangle$
should therefore be understood as the limiting state obtained from
the balanced branch as $T\rightarrow0^{+}$, rather than as a uniquely
selected state of the fully degenerate endpoint.

The two quantities therefore refer to distinct regimes of the same
balanced branch. The residual entropy density $s^{(L)}_{c}$ is a
zero-temperature quantity, defined by the ordered limit
$T\to0^{+}$ after imposing $\delta_L=0$ at finite connectivity. By
contrast, $|\psi^{(L)}_{0,*}\rangle$ is the unique balanced dominant
state, in the symmetrized representation, at finite temperature, $T>0$, whenever
$\omega_L>0$. Its continuation toward $T=0$ defines a limiting
coexistence state, but the fully degenerate endpoint itself does not
select a unique eigenvector. In this sense, $s^{(L)}_{c}$ and
$|\psi^{(L)}_{0,*}\rangle$ are complementary eigenvalue and
eigenvector manifestations, respectively, of the same spectral balance
condition.

\end{document}